\documentclass[11pt,letter,prb,aps,superscriptaddress]{revtex4-1}
\usepackage[utf8x]{inputenc}
\usepackage{amsmath,amssymb}
\usepackage{graphicx}
\usepackage{dcolumn}
\usepackage{bm}
\usepackage{subfigure}
\usepackage{enumerate}
\usepackage{multirow}
\usepackage{tabularx}
\usepackage{threeparttable}
\usepackage{epstopdf}
\usepackage{paralist}
\usepackage{longtable}
\usepackage{color}
\usepackage{soul}
\usepackage{titlesec}
\usepackage{url}
\sethlcolor{yellow}

\newcommand{\ket}[1]{\left| #1 \right\rangle}
\newcommand{\bra}[1]{\left\langle #1 \right|}

\newcommand{\bv}[1]{\mathbf{#1}}

\newcommand{\COMMENT}[1]{}
\newcommand{\textapprox}{{\raise.17ex\hbox{$\scriptstyle\mathtt{\sim}$}}}
\begin{document}


 \title{Frequency-dependent dielectric function of semiconductors with 
application to physisorption}
 \author{Fan Zheng}
 \affiliation{The Makineni Theoretical Laboratories, Department of 
Chemistry, University of Pennsylvania, Philadelphia, PA 19104-6323, USA}
 \author{Jianmin Tao}
\altaffiliation{Email: jianmint@sas.upenn.edu \\
URL: \url{http://www.sas.upenn.edu/~jianmint/}}
 \affiliation{Department of Physics, Temple University, Philadelphia, 
PA 19122-1801, USA}
 \author{Andrew M. Rappe}
 \affiliation{The Makineni Theoretical Laboratories, Department of 
Chemistry, University of Pennsylvania, Philadelphia, PA 19104-6323, USA}

 \begin{abstract}
The dielectric function is one of the most important quantities that 
describes the electrical and optical properties of solids. Accurate 
modeling of the frequency-dependent dielectric function has great significance 
in the study of the long-range van der Waals (vdW) interaction for solids
and adsorption. In this work, we calculate the frequency-dependent 
dielectric functions of semiconductors and insulators using the $GW$ 
method with and without exciton effects, as well as efficient semilocal 
density functional theory (DFT), and compare these calculations with a
model frequency-dependent dielectric function. We find that for 
semiconductors with moderate band gaps, the model dielectric functions,
$GW$ values, and DFT calculations all agree well with each other. However,
for insulators with strong exciton effects, the model dielectric functions
have a better agreement with accurate $GW$ values than the DFT calculations, 
particularly in high-frequency region. To understand this, we repeat the DFT
calculations with scissors correction, by shifting DFT Kohn-Sham energy gap 
to match the experimental band gap. We find that scissors 
correction only moderately improves the DFT dielectric function in 
low-frequency region. Based on the dielectric functions calculated with 
different methods, we make a comparative study by applying these dielectric 
functions to calculate the vdW coefficients ($C_3$ and $C_5$) for 
adsorption of rare-gas atoms on a variety of surfaces. We find that the vdW coefficients obtained with the 
nearly-free electron gas-based model dielectric function agree quite well 
with those obtained from the $GW$ dielectric function, in particular for 
adsorption on semiconductors, leading to an overall error of less than 
7\% for $C_3$ and 5\% for $C_5$. This demonstrates the reliability of
the model dielectric function for the study of physisorption. 
\end{abstract}

 \maketitle 
 \newpage

   \section{Introduction}
    The frequency-dependent dielectric response function, as the linear-order 
response to electric field, plays a central role in the study of the 
electrical and optical properties of solids. It is related to many properties 
of materials. In particular, the static dielectric function has been used in 
the construction of density functional approximations~\cite{Marques11p035119,Skone14p195112}
for the exchange-correlation energy. The frequency-dependent dielectric 
function provides important screening for the van der Waals interaction (vdW) in solids,
because it has been used as an ingredient in the calculation of vdW interactions for 
physisorption and layered materials~\cite{Geim13p419}, which has been one of 
the most interesting topics in condensed matter physics. However, calculation 
of this quantity presents a great challenge to semilocal density functional 
theory (DFT)~\cite{Perdew96p3865,Tao03p146401}, the most popular electronic 
structure method. A fundamental reason is that, while DFT can describe the 
ground-state properties well, it tends to underestimate excitation energies and the 
band gap, due to the absence of electronic nonlocality. For example, the 
widely-used local spin-density approximation (LSDA) and the 
generalized-gradient approximation (GGA) lack the electron-hole interaction 
information for excitons and the discontinuity of energy derivative with respect 
to the number of electrons~\cite{Perdew83p1884,Sham83p1888,Janak78p7165,
Perdew82p1691}. The $GW$ approximation~\cite{Hedin65pA796} for the electron 
self-energy provides a highly-accurate method for describing the 
single-particle spectra of electrons and holes. It yields accurate fundamental 
band gaps of solids~\cite{Zhu91p14142,Onida02p601}. Based on the $GW$ 
approximation, the Bethe-Salpeter equation (BSE) can be solved to capture 
electron-hole interactions~\cite{Onida95p818,Rohlfing98p2312}. Therefore, 
$GW$+BSE has been widely used to calculate optical spectra and light 
absorption, and the results are used as references for other 
methods~\cite{Schleife09p035112,Hsueh11p085404,Qiu13p216805}. However, 
as a cost of high accuracy, this method is computationally demanding, and 
thus it is not practical for large systems. As such, accurate modeling of 
the dielectric functions of semiconductors and insulators with a simple 
analytic function of frequency is highly desired.

Many model dielectric functions have been proposed~\cite{Penn62p2093,Penn87p35,Levine82p6310,Lines90p3372,Kim92p11749}. Most of them have been devoted to the static limit, while the study of the frequency-dependent 
dielectric function is quite limited. Based on a picture of the nearly-free 
electron gas, Penn derived a simple model dielectric function. This model 
was modified by Breckenridge, Shaw, and Sher to satisfy the Kramers-Kronig
relation~\cite{Breckenridge74p2483}. The modified Penn model has been used 
to calculate the vdW coefficient $C_3$ for the  adsorption of atoms on 
surfaces~\cite{Vidali81pL374} and the dielectric screening effect for the 
vdW interaction in solids~\cite{Tao15p164302}. In particular, 
Tao and Rappe~\cite{Tao14p106101} have recently applied the frequency-dependent 
model dielectric function and a simple yet accurate model dynamic multipole 
polarizability to calculate the leading-order as well as higher-order vdW 
coefficients $C_3$ and $C_5$ for atoms on a variety of solid surfaces. The 
results are consistently accurate.

To have a better understanding of this model dielectric function, in the 
present work, we perform $GW$ quasiparticle calculation, by solving BSE, 
aiming to provide a robust reference for benchmarking the model 
frequency-dependent dielectric function. To achieve this goal, we compare 
the model dielectric functions with the high-level $GW$ calculations for 
several typical semiconductors and insulators: silicon, diamond, GaAs, LiF, 
NaF and MgO. As an interesting comparison, we also calculate the dielectric function with the GGA
exchange-correlation functional~\cite{Perdew96p3865}. Based on these dielelctric 
calculations, the vdW coefficients on the various surfaces are also calculated 
and compared to reference values.
To have a better understanding of the performance of DFT, we repeat our
DFT dielectric function calculation after shifting the Kohn-Sham eigen-energies to match experimental 
band gaps (scissors correction)~\cite{Levine89p1719}.
      
\section{Computational Details}
 \subsection{Model dielectric function}
The Penn model is perhaps the most widely-used model dielectric function for 
semiconductors. It was derived from the nearly-free electron gas. However, 
this model violates the Kramers-Kronig relation~\cite{Penn62p2093}. To fix 
this problem, Breckenridge, Shaw, and Sher~\cite{Breckenridge74p2483} 
proposed a modification, in which the imaginary part takes the expression

 \begin{flalign}
  \epsilon_2\left(\omega\right) & = \pi \bar{\omega}^2_p\left[\omega_g-\Delta\left(\omega^2-\omega_g^2\right)^{1/2}\right]^2\Big/\left[2\omega^3\left(\omega^2-\omega_g^2\right)^{1/2}\right]. 
 \end{flalign}

\noindent Here, $\omega$ is a real frequency within the range 
$\omega_g \le \omega \le 4\epsilon_F\sqrt{1+\Delta^2}$\ ~\cite{Tao14p106101,
Vidali81pL374}, $\epsilon_F=(3\pi^2 {\bar n})^{2/3}/2$ is the Fermi energy, 
and $\bar n$ is the average valence electron density of the bulk solid. 
$\Delta = \omega_g/4\epsilon_F$, and $\omega_g$ is the effective energy 
gap, which can be determined from optical dielectric constant 
$\epsilon_1\left(0\right)$ by solving the Penn's model:

\begin{flalign}\label{eq:penn}
\epsilon_{1}\left(0\right) & = 
1 + \left(\omega_p^2/\omega_g^2\right)\left(1-\Delta\right).
\end{flalign}

Here, we use this expression to calculate $\omega_g$ from the 
experimental static dielectric constant 
for diamond, LiF, NaF, and MgO. (In Ref.~\citenum{Tao14p106101}, 
the \textit{ab initio} values of $\epsilon_1(0)$, 
rather than experimental values, were used. Since the two sets of 
values are very close to each other, it does not make a noticeable difference.) 
For other materials, $\epsilon_1(0)$ values are taken from 
the literatures~\cite{Vechten69p891,Breckenridge74p2483}. The real part of the 
dielectric function can be obtained from the Kramers-Kronig' relation: 
$\epsilon_1(\omega)=1+\frac{1}{\pi}\mathcal{P}\int_{-\infty}^{+\infty}\epsilon_2\left(\omega'\right)/\left(\omega'-\omega\right)d\omega'$. The result is
given by~\cite{Tao15p164302}
\begin{flalign}\label{model}
\epsilon_1(iu) & = 1+\frac{{\bar\omega}_p^2}{u^2}\bigg[
\frac{(1-\Delta^2)y}{P}-
\frac{\omega_g^2-(\omega_g^2+u^2)\Delta^2}{2u\sqrt{\omega_g^2+u^2}}
{\rm ln}\frac{I_{+}}{I_{-}}\bigg]  \nonumber\\
 & + \frac{2{\bar\omega}_p^2\Delta}{u^2}\bigg\{\frac{\omega_g}{u}\bigg[
{\rm tan}^{-1}\bigg(\frac{{\omega}_g P}{u}\bigg)-
{\rm tan}^{-1}\bigg(\frac{{\omega}_g}{u}\bigg)\bigg]
+\frac{1}{P}-1\bigg\},
\end{flalign}
where $I_{\pm} = [(1+y^2)(1+u^2/\omega_g^2)]^{1/2} \pm uy/\omega_g$,
$y = 1/\Delta$, and $P=(1+y^2)^{1/2}$. Vidali and Cole~\cite{Vidali81pL374}
found that this model dielectric function agrees well with experimental
values GaAs~\cite{Philipp62p92,Philipp63p1550,Sturge62p768,QueueSystems}.

\subsection{DFT calculations}
The DFT calculation of the dielectric function for solids was performed with
the plane-wave density functional theory (DFT) package 
QUANTUM-ESPRESSO~\cite{Giannozzi09p395502}, with the GGA 
exchange-correlation functional~\cite{Perdew96p3865}. The 
norm-conserving, designed non-local pseudopotentials were generated with the OPIUM 
package~\cite{Rappe90p1227,Ramer99p12471}. With the single-particle 
approximation, the imaginary part of the dielectric response function in the 
long-wavelength limit can be expressed as~\eqref{eq:dielectric}
  
      \begin{flalign}\label{eq:dielectric}
      \epsilon_{2,j}(\omega) &= \frac{\pi}{2\epsilon_0}\frac{e^2}{m^2\left(2\pi\right)^4\hbar\omega^2}\sum_{c,v}\int_{{\rm BZ}} d\bv{k} \left| \bra{c,\bv{k}}p_j\ket{v,\bv{k}}\right|^2 \delta(\omega_{c,\bv{k}}-\omega_{v,\bv{k}}-\omega) 
      \end{flalign}
  
  \noindent In this equation, $c$ and $v$ represent the conduction and valence 
bands with eigen-energy $\hbar\omega_{n}$, and $\bv{k}$ is the Bloch 
wave vector. In Cartesian coordinates, $j$ indicates $x$, $y$ or $z$. In 
practice, the real part of the dielectric function, $\epsilon_1\left(i u\right)$ 
expressed in terms of the imaginary frequency $iu$, can be obtained from the 
imaginary part via the Kramers-Kronig relation. 

It is well known that semilocal DFT tends to underestimate the band gaps of 
semiconductors and insulators. To understand the role of band gap, we 
repeated the DFT calculation, replacing the Kohn-Sham HOMO-LUMO energy gap 
with the experimental band gap\cite{Levine89p1719}. This scissor correction will allow us to
study the band gap effect on the dielectric function~\cite{Nastos05p045223} by
\begin{flalign} 
\omega_{mn}=\omega_{mn}^{\rm LDA}+\Delta\omega,
\end{flalign}
where $\omega_{mn}$ is the energy difference between bands $m$ and $n$, 
and $\Delta\omega$ is the scissor correction for reproducing the experimental band 
gap. In this work, this correction is applied to the insulators via the 
rigid shifting of the imaginary part of the dielectric functions.

\subsection{$GW$ and BSE calculations}
The $GW$ calculations including electron-electron screening are carried out 
using the BerkeleyGW package~\cite{Hybertsen86p5390,Rohlfing00p4927,Deslippe12p1269}. 
In the $GW$ approximation, the quasiparticle energy is given by

\begin{flalign}\label{eq:self}
E_{n\bv{k}}^{{\rm QP}} = E_{n\bv{k}}^{{\rm MF}} + \bra{\psi_{n,\bv{k}}}\Sigma\left(E\right) - V_{{\rm XC}}\ket{\psi_{n,\bv{k}}}
\end{flalign}

\noindent where $\Sigma$ is the self-energy and $\psi_{n\bv{k}}$ is a mean-field 
wave function. $V_{{\rm XC}}$ is the exchange-correlation potential obtained 
from the GGA or LDA functionals. The mean-field part of the DFT electronic 
structure calculations was performed with QUANTUM-ESPRESSO. First, the static 
dielectric matrix $\epsilon\left(\bv{q};0\right)$ within the random-phase 
approximation (RPA) is calculated. Then, the generalized plasmon-pole 
and static coulomb hole and screened exchange approximation (COHSEX) were used to evaluate the 
self-energy $\Sigma$. In order to have accurate quasiparticle energies, 
the convergence of band energies with number of empty bands in the dielectric 
matrix and Coulomb hole (COH) self-energy evaluations, and the convergence 
versus plane-wave cutoff were carefully tested~\cite{Malone13p105503}. Due to the 
significance of electron-hole interaction in determining the optical 
response, the BSE was solved to reveal the effect of excitons on light absorption. 
This is particularly important for ionic solids, such as LiF, NaF, and MgO, 
with strongly bound excitons. To perform BSE calculations, the electron-hole 
kernel terms evaluated on a coarse $k$ point grid were interpolated onto a dense 
grid. By diagonalizing the kernel matrix, exciton eigenvalues $\Omega^S$ and 
eigenfunctions $\ket{S}$ were solved and used in the calculation of the optical 
dielectric function~\cite{Rohlfing00p4927}:

\begin{flalign}\label{eq:bse}
\epsilon_2\left(\omega\right) = \frac{16\pi^2 e^2}{\omega^2}\sum_{S}\left|\bv{e}\cdot\bra{0}\bv{v}\ket{S}\right|^2\delta\left(\omega-\omega^S\right)
\end{flalign}

\noindent where $S$ is the exciton state with exciton energy $\omega^S$. The 
dielectric function with imaginary frequency dependence can be easily obtained.

\subsection{vdW coefficients}
The vdW interaction is crucial for adsorption of atoms or molecules on solid 
surfaces, while adsorption on solids is fundamentally important in probing 
the surface structures and properties of bulk solids (e.g., atomic or molecular 
beam scattering) as well as catalysis and hydrogen storage (e.g., surface 
adsorption on fullerenes, nanotubes and graphene). In the process of 
physisorption, the instantaneous multipole due to the electronic charge 
fluctuations of a solid will interact with the dipole, quadrupole and octupole
moments of adsorbed atoms or molecules, giving rise to vdW attraction. 
However, semilocal DFT often fails to describe this process, because the
long-range vdW interaction is missing in semilocal DFT. Many
attempts~\cite{Dion04p246401,Granatier11p3743,Klimes11p195131,Lazic05p245407,Chen12p424211,Grimme04p1463,Becke07p154108,Tkatchenko09p073005,Silvestrelli08p053002,Tkatchenko13p074106,Tao10p233102,Tao12p18,Tao14p141101,Tao15p164302,Tao16p031102} 
have been made 
to capture this long-range part, such as nonlocal vdW-DF 
functional~\cite{Dion04p246401} and density functional dispersion 
correction~\cite{Grimme10p154104,Grimme06p1787}. It has been shown that
with a proper dispersion correction, the performance of ordinary DFT 
methods can be significantly improved~\cite{Tao14p106101}. This combined 
DFT+vdW method has been widely used in electronic structure calculations 
of molecules and solids~\cite{Ruiz12p146103,Ma11p033402,Fang12p10692,Sirtl13p11054,Tao12p18}.

The vdW coefficients for adsorption on solid surfaces were calculated in terms 
of the dielectric function and the dynamic multipole polarizability. The molecular dynamic 
multipole polarizability was computed from a simple yet accurate model described 
in Refs.~\citenum{Tao12p18,Tao10p233102}. 
The molecular electronic charge density was obtained from Hartree-Fock calculations using 
GAMESS~\cite{Schmidt93p1347,Dykstra11p1167}. With the imaginary frequency 
dependent dielectric function and the atomic polarizabilities, the vdW 
coefficients $C_3$ and $C_5$ were calculated 
from~\cite{Dalvit11,Zaremba76p2270,Tao14p106101}
  
  \begin{flalign}\label{eq:vdw_coefficient}
       C_{2l+1} = \frac{1}{4\pi}\int_0^{\infty}du \alpha_l\left(iu\right)\frac{\epsilon_1\left(iu\right) -1 }{\epsilon_1\left(iu\right) +1 }
  \end{flalign}
  
  \noindent where $l=1$ describes the interaction of the instantaneous dipole 
moment of an atom with the surface, while $l=2$ describes the interaction of 
the quadrupole moment of the atom with the surface. $\epsilon_1$ is the 
real part of the dielectric function of the bulk solid, and $\alpha_l(iu)$ is the dynamic multipole polarizability.

\section{Results and discussion}
\subsection{Dielectric function}
The experimental values of the frequency dependent dielectric 
function are not directly available in the literature, but they can be 
extracted from experimental optical data~\cite{Vidali81pL374}. On the 
other hand, comparison of the calculated static dielectric function to 
experiment is indicative of the accuracy of the calculated frequency 
dependence.

Table~\ref{basic_values} shows the calculated and experimental static
dielectric functions of several semiconductors and insulators. The effective 
energy gaps derived from the static dielectric functions are also listed in 
Table~\ref{basic_values}. From Table~\ref{basic_values}, we can observe 
that the $GW$+BSE static dielectric functions agree very well with 
experiments for all the materials considered, while the $GW$ values 
have better agreement with experiments for semiconductors than for insulators,
due to the strong exciton effect in insulators~\cite{Yang15p035202}. 
Table~\ref{basic_values} also shows that DFT tends to overestimate the static dielectric function, in particular for insulators. This overestimate was also observed in the adiabatic local density approximation within the time-dependent DFT formalism~\cite{Aulbur1999,vanFaassen02p186401,vanFaassen03p1044}.
However, as shown in Table~\ref{basic_values}, a scissors correction cannot cure this overestimate tendency problem. We attribute this problem to the lack of electronic nonlocality of semilocal DFT. The frequency-dependent dielectric function for each material is discussed below.

\begin{figure}[h]
 \includegraphics[width=4.5in]{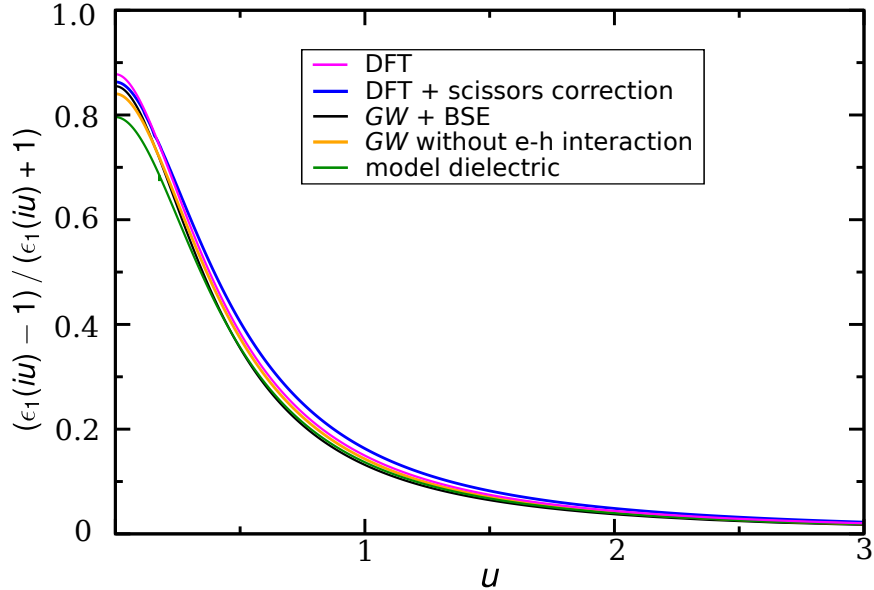}
  \caption{$\left(\epsilon_1\left(iu\right)-1\right)/\left(\epsilon_1\left(iu\right)+1\right)$ of {\bf silicon} with respect to frequency $u$ (in hartree) 
calculated from DFT, DFT+scissors correction, $GW$, $GW$+BSE and model dielectric 
function.}\label{si}
\end{figure}

\centerline{\bf Silicon}
 Fig.~\ref{si} shows $\left(\epsilon_1\left(iu\right)-1\right)/\left(\epsilon_1\left(iu\right)+1\right)$ of Si semiconductor calculated with the DFT-GGA, DFT+scissor correction, $GW$, $GW$+BSE and the model dielectric function of Eq.~(\ref{model}). The DFT calculated band gap is 0.62 eV, which significantly underestimates the experimental band gap by 0.55 eV. The experimental static dielectric constant is 11.7, which is reproduced by $GW$+BSE calculations (Table.~\ref{basic_values}). From Fig.~\ref{si}, DFT gives quite accurate description of optical response in terms of $\left(\epsilon_1-1\right)/\left(\epsilon_1+1\right)$, although it gives slightly higher dielectric constant than $GW$+BSE at zero frequency. At low frequencies, the model dielectric function underestimates the $GW$ value. This underestimate is due to the error in the effective energy gap $\omega_g$~\cite{Breckenridge74p2483}, which is slightly overestimated. Nevertheless, the model dielectric function agrees with $GW$+BSE results quite well, particularly in the high-frequency region.

  \begin{table*}[h]
  \caption{Experimental band gaps (fundamental), DFT scissors band gap 
corrections ($\Delta_{\rm corr}=E_g^{\rm expt}-E_g^{\rm DFT}$), effective 
energy gaps ($\omega_g$) of the model dielectric function, and dielectric 
constants ($\epsilon_{0}$) of the model dielectric function, DFT and 
$GW$+BSE.}\label{basic_values}
  \begin{tabular}{ccccccc}\hline\hline
                       &  Si   & GaAs&  C   & LiF   & NaF & MgO \cr\hline
$E_g^{\rm expt}$ (eV) & 1.17$^b$  &1.52$^b$ & 5.48$^b$ &14.20$^b$  &11.70$^d$& 7.83$^b$ \cr 
$\Delta_{\rm corr}$ (eV)     & 0.49  &1.12 & 1.21 & 5.20  & 5.58&  3.27 \cr
$\omega_g$ (eV)      &4.8$^a$ &4.3$^b$ & 13.0$^c$ &23.3$^c$& 20.5$^c$& 15.5$^c$ \cr
$\epsilon_{0}^{\rm expt}$   & 12.0$^b$  &11.3$^b$ & 5.9$^b$  & 1.9$^b$   & 1.7$^e$ & 3.0$^b$  \cr
$\epsilon_{0}^{\rm model}$   &  9.8  &8.9  & 4.4  & 1.6   & 1.5 & 2.3  \cr 
$\epsilon_{0}^{\rm DFT}$     & 15.4  &11.0 & 6.6  & 2.5   & 2.3 &  4.1 \cr 
$\epsilon_{0}^{\rm DFT+sci.}$& 13.6  &8.1  & 5.7  & 2.1   & 1.9 &  3.5 \cr 
$\epsilon_{0}^{GW}$          &11.5   &10.7  &5.1  &1.8  & 1.6 & 2.6      \cr
$\epsilon_{0}^{GW+{\rm BSE}}$& 12.7  &11.0 & 5.7  &1.9    & 1.7 & 2.9  \cr   \hline \hline
  $^a$ Ref.~\cite{Breckenridge74p2483}\\
  $^b$ Ref.~\cite{Vechten69p891}\\
  $^c$ Obtained from Eq.~\eqref{eq:penn}\\
  $^d$ Ref.~\cite{Poole75p5179}\\
  $^e$ Ref.~\cite{Lines90p3372}\\
  \end{tabular}
  \end{table*}

\begin{figure}
 \includegraphics[width=4.5in]{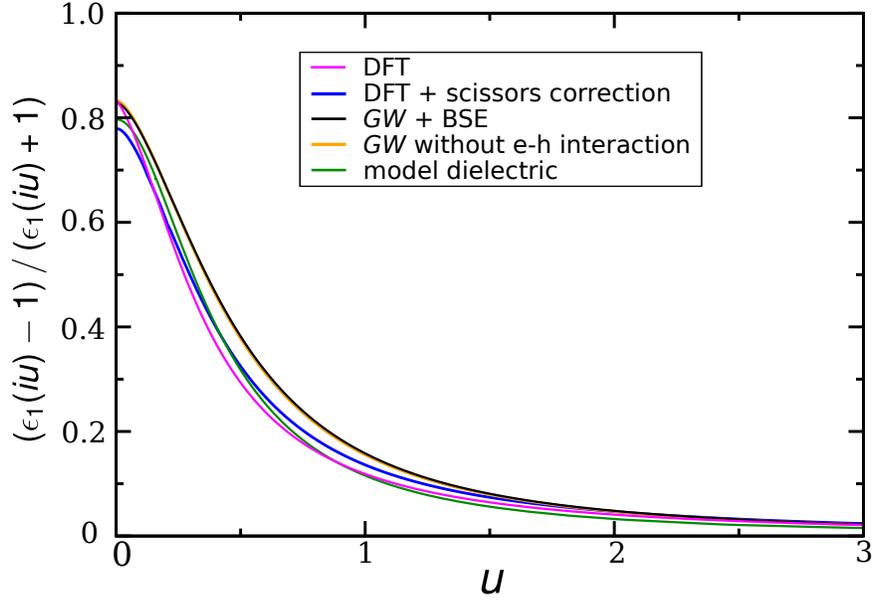}
 \caption{$\left(\epsilon_1\left(iu\right)-1\right)/\left(\epsilon_1\left(iu\right)+1\right)$ of {\bf GaAs} with respect to  frequency $u$ (in hartree) calculated 
from DFT, DFT+scissors correction, $GW$, $GW$+BSE, and model dielectric 
function.}\label{gaas}
\end{figure}

\centerline{\bf GaAs}
 Fig.~\ref{gaas} shows the computed dielectric functions of GaAs. $GW$ and $GW$+BSE show very similar dielectric functions, indicating the weak exciton effect in GaAs\cite{Nam76p761}, and strong dielectric screening effect. DFT and model dielectric functions slightly underestimate $GW$+BSE values, which is because of the higher absorption calculated with $GW$ and $GW$+BSE than that with DFT. In general, similar to silicon, all the methods yield dielectric functions close to each other, in particular in the high-frequency region. This similarity is largely due to the fact that both semiconductors have similar band gaps and dielectric constants, as shown in
Table~\ref{basic_values}. 

\begin{figure}
 \includegraphics[width=4.5in]{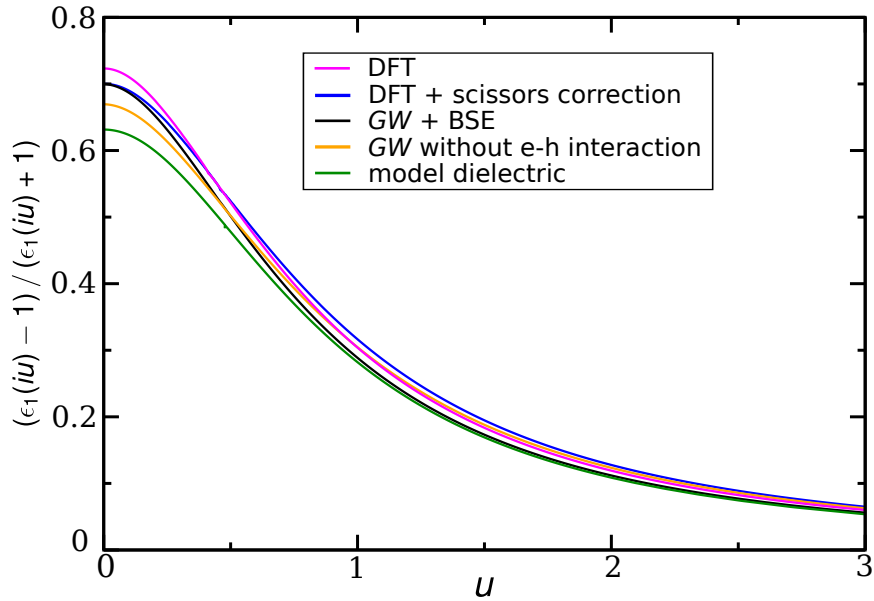}
  \caption{$\left(\epsilon_1\left(iu\right)-1\right)/\left(\epsilon_1\left(iu\right)+1\right)$ of {\bf diamond} with respect to frequency $u$ (in hartree) 
calculated from DFT, DFT+scissors correction, $GW$, $GW$+BSE, and model 
dielectric function.}\label{diamond}
\end{figure}

\centerline{\bf Diamond}
 The dielectric function of diamond is shown in Fig.~\ref{diamond}. Diamond 
shares similar geometric and electronic structures with silicon, but with much 
 larger band gap. In this case, the overestimation of dielectric 
function from DFT and the underestimation from model dielectric function are 
more pronounced than those for silicon at low frequencies. This difference is 
mainly due to the discrepancy between the Penn model effective band gap (slightly overestimated) and the $GW$ or $GW$+BSE value. However, as energy increases 
to the high-energy region, this discrepancy vanishes, matching the model dielectric 
function to $GW$+BSE results very well. 

\begin{figure}
 \includegraphics[width=4.5in]{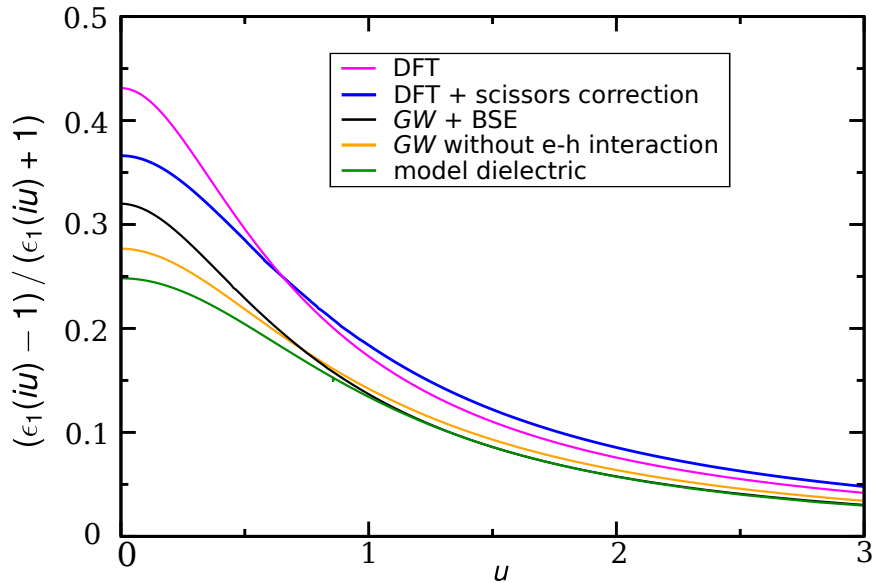}
 \caption{$\left(\epsilon_1\left(iu\right)-1\right)/\left(\epsilon_1\left(iu\right)+1\right)$ of {\bf LiF} 
with respect to frequency $u$ (in hartree) calculated from DFT, DFT+scissors 
correction, $GW$, $GW$+BSE, and model dielectric function.}\label{lif}
\end{figure}

\centerline{\bf LiF}
 LiF is a prototypical material with strong exciton effect on its optical 
absorption~\cite{Abbamonte08p12159}. As shown in Fig.~\ref{lif}, at low energies, 
$GW$+BSE including electron-hole interaction yields higher value compared to 
the dielectric function without electron-hole interaction, which corresponds to 
the exciton absorption. Due to the same discrepancy observed in diamond, the 
model dielectric function underestimates the response near zero energy, but 
matches $GW$-BSE result well in the high-energy region. The vdW coefficients measure 
the strength of the dielectric response of a bulk solid to the instantaneously 
induced multipole moment of the adsorbed atom or molecule. They are integrated over 
the whole energy range, including both low-energy and high-energy dielectric contributions. Thus, 
the noticeable discrepancy observed in the low-energy part 
has minor effect on the overall vdW coefficients. However, the DFT-calculated 
dielectric response seriously overestimates the response in the whole energy 
spectrum, compared to $GW$+BSE, leading to significantly overestimated 
vdW coefficients, as shown in the Table~\ref{si_vdw}. 
This overestimation problem cannot be fixed even with scissors 
correction to the DFT band gap. Comparison of $GW$-BSE with $GW$ (without 
electron-hole interaction) suggests that there is an important exciton effect 
on the dielectric function in the low-energy range. This suggests that
semilocal DFT may not fully capture this exciton effect as well as the many-body effect. As a result, semilocal DFT tends to overestimate the dielectric function, although it slightly underestimates the dielectric function for semiconductors.

\centerline{\bf NaF}
 NaF is another prototypical material with strong exciton effects. Figure~\ref{lif} 
shows the comparison of the dielectric function evaluated with all the methods
discussed above. From Fig~\ref{lif}, we observe that the model dielectric function 
still underestimates the response near zero frequency, but with overall good quality 
matching of $GW$+BSE results. However, semilocal DFT and scissors-corrected 
semilocal DFT strongly overestimate the dielectric function magnitude for the whole 
frequency range, reflecting the inadequacy of semilocal DFT, as observed in other
ionic solids. 

\begin{figure}
 \includegraphics[width=4.5in]{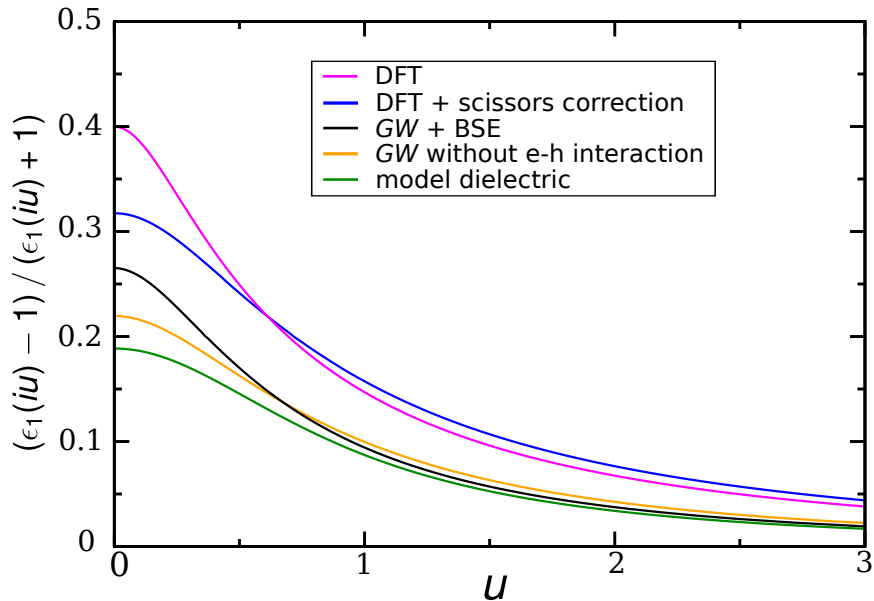}
 \caption{$\left(\epsilon_1\left(iu\right)-1\right)/\left(\epsilon_1\left(iu\right)+1\right)$ of {\bf NaF} with respect to frequency $u$  (in hartree) calculated from 
DFT, DFT+scissors correction, $GW$, $GW$+BSE, and model dielectric function.}
\label{naf}
\end{figure}

\begin{figure}
 \includegraphics[width=4.5in]{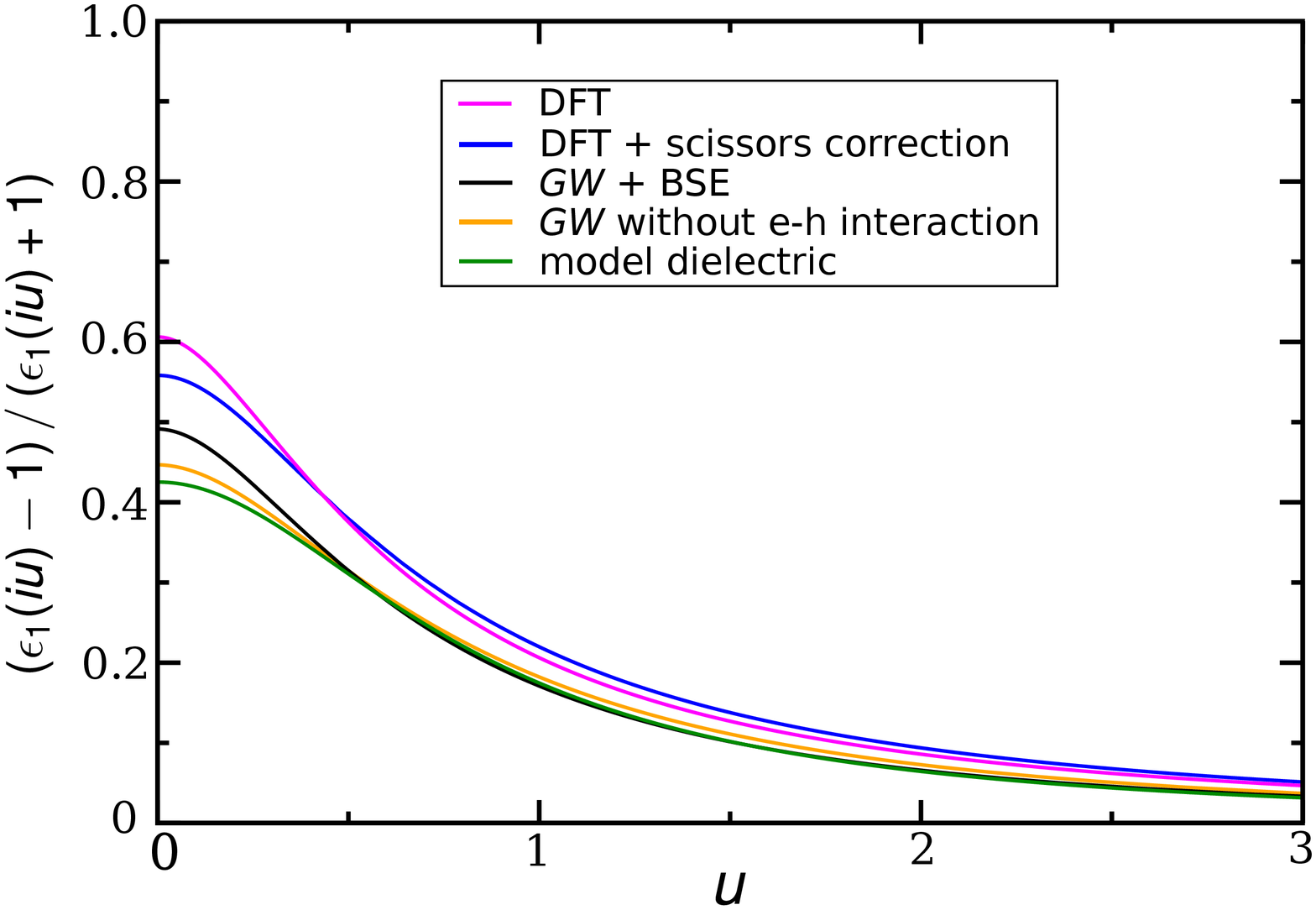}
 \caption{$\left(\epsilon_1\left(iu\right)-1\right)/\left(\epsilon_1\left(iu\right)+1\right)$ of {\bf MgO} with respect to frequency $u$ (in hartree) calculated from 
DFT, DFT+scissors correction, $GW$, $GW$+BSE, and model dielectric function.}
\label{mgo}
\end{figure}

\centerline{\bf MgO}
As a support for variety of catalytic reactions~\cite{Zhang07p2228,Yoon05p403},  
MgO has attracted great attension in recent
years. Accurate calculation of the dielectric function for the vdW interaction is 
significantly important for the prediction of the correct chemical reaction path 
and energy barrier. As shown in Fig.~\ref{mgo}, MgO also shows strong exciton 
effect, leading to obvious but less pronounced deviation of the DFT curve from the 
$GW$+BSE calculation, compared to other ionic solids considered here. On the other
hand, the model dielectric function agrees with $GW$+BSE values rather well. \\[1in]

\subsection{vdW Coefficients for adsorption on surfaces of solids}
The vdW coefficients $C_3$ and $C_5$ can be calculated from
Eq.~(\ref{eq:vdw_coefficient}) with the model dynamic multipole polarizability 
given by~\cite{Tao12p18} 
\begin{eqnarray}\label{multipole}
\alpha_l(iu) = 
\frac{2l+1}{4\pi d_l}\int_0^{R_l} 
dr~4\pi r^2 \frac{r^{2l-2}~d_l^4\omega_l^2}{d_l^4\omega_l^2+u^2},
\end{eqnarray}
where $R_l$ is the effective vdW radius and $d_l$ is a parameter
introduced to satisfy the exact zero- and high-frequency limits. Numerical
tests show that the model can generate vdW coefficients for diverse atom pairs 
in excellent agreement with accurate reference values, with mean absolute 
relative error of only $3\%$. To benchmark our model dielectric function 
for adsorption, we calculate the vdW coefficients with several dielectric 
functions obtained from $GW$, $GW$+BSE and DFT-GGA methods, and compare them 
to the vdW coefficients obtained from the model dielectric function and 
accurate reference values. The results are shown in Table~\ref{si_vdw}.  

From Table~\ref{si_vdw}, we observe that the vdW coefficients calculated 
from the model dielectric function are close to the reference values. They agree 
quite well with the $GW$ and $GW$+BSE values, with mean absolute relative 
deviations of 2\% for $C_3$ and 5\% $C_5$ from those calculated with the $GW$ 
dielectric function, and 4\% for $C_3$ and 8\% for $C_5$ from those evaluated 
with the $GW$+BSE dielectric function, 
respectively. The strong exciton observed in ionic solids LiF, NaF and MgO has
some effect on the vdW coefficients. But this effect is relatively small for
the vdW coefficients evaluated with $GW$ and $GW$+BSE dielectric function, 
as the dielectric enhancement by exitons only appears within small frequency 
range. The model dielectric function can also accounts for exitons via
the static dielectric function part, the vdW coefficients evaluated from the
model dielectric function agree reasonably well with these two {\em ab initio}
values even for materials with strong exiton effect, as found in the ionic 
solids considered here. However, we find that the DFT-GGA significantly 
overestimates vdW coefficients by 30\% for $C_3$ and 33\% for $C_5$, due to 
the overestimation of the dielectric functions in the whole frequency range. 
Moreover, scissors correction to the DFT dielectric function shows little 
improvement of vdW coefficient. The detail of DFT calculations can be found 
from Table~\ref{si_vdw}.

  \begin{longtable}{c|cc|cc|cc|cc|cc|cc}
  \caption{vdW coefficients (in a.u.) between rare gas atoms and the surfaces of semiconductors and insulators. These are calculated by DFT, DFT+ scissors, $GW$, $GW$+BSE, and model dielectric. The reference values of He atom on all surfaces are from Ref.~\citenum{Vidali81pL374}. Values for other atoms are from Ref.~\citenum{Vidali81p10}. MRE = mean relative error. MARE = mean absolute relative error.}\label{si_vdw}\\
  \hline\hline
\multicolumn{1}{c|} {   }     &  \multicolumn{2}{c|} {DFT}   &  \multicolumn{2}{c|} {DFT+sci.}   &  \multicolumn{2}{c|} {$GW$}   & \multicolumn{2}{c|} {$GW$+BSE}   & \multicolumn{2}{c|} {Model diele.}   & \multicolumn{2}{c} {Reference} \endfirsthead
\caption{Continued.}\\
\hline
\endhead
\hline
{\bf Silicon} &   $C_3$ &  $C_5$ & $C_3$ & $C_5$  &  $C_3$ & $C_5$ & $C_3$ & $C_5$ & $C_3$ & $C_5$ & $C_3$ & $C_5$ \cr 
H    & 0.105 & 0.416 & 0.107 & 0.425 & 0.100  & 0.395  & 0.101    &  0.402  &  0.096    &  0.383 & 0.102  &   0.366 \cr
He   & 0.046 & 0.083 & 0.047 & 0.086 & 0.043  & 0.078  & 0.044    &  0.080  &  0.042    &  0.076 & 0.042  &   0.076 \cr
Ne   & 0.096 & 0.262 & 0.099 & 0.270 &  0.090  &    0.246  &    0.093    &  0.253  &  0.088    &  0.241 & 0.089  &    0.241    \cr
Ar   & 0.330 & 1.632 & 0.338 & 1.676 &  0.312  &    1.541  &    0.319    &  1.578  &  0.304    &  1.502 & 0.310  &    1.490    \cr
Kr   & 0.468 & 2.888 & 0.479 & 2.959 &  0.443  &    2.735  &    0.452    &  2.794  &  0.431    &  2.659 & 0.449  &    2.644    \cr
Xe   & 0.802 & 6.613 & 0.822 & 6.782 &  0.758  &    6.254  &    0.775    &  6.395  &  0.738    &  6.088 & 0.655  &    5.469    \cr\hline
{\bf GaAs}  &   $C_3$ &  $C_5$ & $C_3$ & $C_5$  &   $C_3$ & $C_5$ &    $C_3$ & $C_5$ &    $C_3$ & $C_5$ &     $C_3$ & $C_5$ \cr 
H    & 0.089 & 0.350 & 0.091 & 0.361 &  0.100  &   0.400   &   0.101   &   0.401  &    0.092   &   0.362  & 0.091   &   0.351 \cr
He   & 0.038 & 0.069 & 0.040 & 0.073 &  0.044  &   0.081   &   0.045   &   0.081  &    0.039   &   0.071  & 0.041   &   0.072 \cr              
Ne   & 0.080 & 0.219 & 0.084 & 0.230 &  0.093  &   0.255   &   0.094   &   0.256  &    0.082   &   0.224  & 0.081   &   0.227 \cr
Ar   & 0.277 & 1.364 & 0.287 & 1.422 &  0.318  &   1.577   &   0.320   &   1.585  &    0.285   &   1.407  & 0.285   &   1.417 \cr
Kr   & 0.393 & 2.420 & 0.407 & 2.513 &  0.451  &   2.785   &   0.453   &   2.797  &    0.406   &   2.500  & 0.412   &   2.523 \cr
Xe   & 0.674 & 5.548 & 0.701 & 5.768 & 0.775  &   6.386   &   0.779   &   6.416  &    0.693   &   5.715  & 0.603   &   5.242 \cr\hline
{\bf Diamond}  &   $C_3$ &  $C_5$ & $C_3$ & $C_5$  &   $C_3$ & $C_5$ &    $C_3$ & $C_5$ &    $C_3$ & $C_5$ &     $C_3$ & $C_5$ \cr 
H    & 0.113 & 0.470 & 0.112 & 0.468 &   0.108   &   0.448  &   0.109  &    0.452 &   0.101  &   0.422  & 0.112   &   0.407 \cr
He   & 0.057 & 0.105 & 0.057 & 0.106 &   0.054   &   0.102  &   0.054  &    0.101 &   0.051  &   0.095  & 0.051   &   0.097 \cr
Ne   & 0.123 & 0.334 & 0.124 & 0.338 &   0.119   &   0.323  &   0.118  &    0.320 &   0.110  &   0.300  & 0.116   &   0.308 \cr
Ar   & 0.390 & 1.961 & 0.061 & 0.257 &   0.374  &    1.882 &    0.374 &     1.881&    0.350  &   1.761  & 0.375  &    1.781 \cr
Kr   & 0.543 & 3.378 & 0.542 & 3.378 &   0.519   &   3.233  &   0.521  &    3.243 &   0.486  &   3.032  & 0.526   &   3.069 \cr
Xe   & 0.960 & 7.857 & 0.963 & 7.871 &   0.922   &   7.534  &   0.921  &    7.539 &   0.861  &   7.047  & 0.737   &   6.132 \cr\hline
{\bf LiF} &   $C_3$ &  $C_5$ & $C_3$ & $C_5$  &   $C_3$ & $C_5$ &    $C_3$ & $C_5$ &    $C_3$ & $C_5$ &     $C_3$ & $C_5$ \cr 
H     & 0.066 & 0.276 & 0.061 & 0.257 &    0.046  &    0.194    &  0.050    &  0.208   &   0.042   &   0.178  &     0.048   &   0.169   \cr
He    & 0.033 & 0.062 & 0.032 & 0.060 &    0.024 &     0.045   &   0.025   &   0.047  &    0.022  &    0.041 &      0.023  &    0.042  \cr              
Ne    & 0.073 & 0.198 & 0.071 & 0.192 &    0.052 &     0.142   &   0.055   &   0.148  &    0.048  &    0.131 &      0.048  &    0.133  \cr
Ar    & 0.229 & 1.153 & 0.218 & 1.097 &    0.163 &     0.821   &   0.173   &   0.868  &    0.150  &    0.756 &      0.155  &    0.756  \cr
Kr    & 0.320 & 1.984 & 0.302 & 1.872 &    0.225 &     1.405   &   0.240   &   1.494  &    0.207  &    1.292 &      0.219  &    1.294  \cr
Xe    & 0.568 & 4.631 & 0.541 & 4.395 &    0.402 &     3.281   &   0.425   &   3.475  &    0.370  &    3.019 &      0.313  &    2.561  \cr\hline       
{\bf NaF} &   $C_3$ &  $C_5$ & $C_3$ & $C_5$  &   $C_3$ & $C_5$ &    $C_3$ & $C_5$ &    $C_3$ & $C_5$ &     $C_3$ & $C_5$ \cr 
H    & 0.059 & 0.241 & 0.052 & 0.220 &    0.035  &    0.146  &    0.039   &   0.160  &    0.035  &    0.147   &  0.038   &   0.137 \cr
He   & 0.029 & 0.054 & 0.027 & 0.052 &    0.018  &    0.033  &    0.019   &   0.035  &    0.018  &    0.033   &  0.018   &   0.033 \cr              
Ne   & 0.064 & 0.172 & 0.061 & 0.165 &    0.039  &    0.105  &    0.041   &   0.111  &    0.039  &    0.105   &  0.037   &   0.104 \cr
Ar   & 0.200 & 1.005 & 0.186 & 0.940 &    0.122  &    0.613  &    0.131   &   0.657  &    0.122  &    0.615   &  0.123   &   0.600 \cr
Kr   & 0.280 & 1.733 & 0.258 & 1.603 &    0.169  &    1.054  &    0.183   &   1.138  &    0.170  &    1.058   &  0.174   &   1.032 \cr
Xe   & 0.495 & 4.040 & 0.463 & 3.764 &    0.300  &    2.454  &    0.322   &   2.638  &    0.301  &    2.462   &  0.248   &   2.059 \cr\hline
{\bf MgO} &   $C_3$ &  $C_5$ & $C_3$ & $C_5$  &   $C_3$ & $C_5$ &    $C_3$ & $C_5$ &    $C_3$ & $C_5$ &     $C_3$ & $C_5$ \cr 
H   & 0.087 & 0.358 & 0.085 & 0.352 &    0.069    &  0.286  &    0.072   &   0.295     & 0.063   &   0.259  & 0.069   &   0.252 \cr
He  & 0.042 & 0.079 & 0.042 & 0.079 &    0.034   &   0.064 &     0.035  &    0.064    &  0.031  &    0.057 &  0.032  &    0.059 \cr              
Ne  & 0.092 & 0.249 & 0.092 & 0.250 &    0.074   &   0.202 &     0.075  &    0.204    &  0.067  &    0.182 &  0.066  &    0.188 \cr              
Ar  & 0.295 & 1.476 & 0.292 & 1.465 &    0.237   &   1.189 &     0.242  &    1.212    &  0.214  &    1.073 &  0.224  &    1.094 \cr              
Kr  & 0.412 & 2.557 & 0.407 & 2.527 &    0.329   &   2.050 &     0.338  &    2.101    &  0.298  &    1.854 &  0.315  &    1.892 \cr              
Xe  & 0.725 & 5.934 & 0.719 & 5.880 &    0.582   &   4.764 &     0.594  &    4.867    &  0.524  &    4.299 &  0.439  &    3.796 \cr\hline
MRE(\%) & 29.3 & 32.4 & 27.0 & 30.4 &7.2&9.7&     10.2  &    12.9    &  1.3  &   3.7 &  -  &    - \cr
MARE(\%) & 30.3 & 33.2 & 27.3 & 30.4 &8.5&9.7&     10.5  &    12.9    &  6.7  &   4.6 &  -  &    - \cr\hline \hline 
  \end{longtable}

   \section{Conclusion}
In summary, we have calculated the frequency-dependent dielectric function
of semiconductors and insulators with the DFT-GGA, $GW$ and $GW$+BSE 
methods. Based on these calculations, we study the accuracy of the modified 
Penn model by comparing the model dielectric function to the highly-accurate 
$GW$ and $GW$+BSE methods. We find that the model dielectric function
agrees quite well with these two methods, in particular for small energy-gap
semiconductors. However, a noticeable discrepancy arises with the increase 
of band gap. A similar trend has been also observed with the DFT-GGA dielectric 
function, which shows even greater disagreement with the $GW$ and $GW$+BSE methods, 
compared to the model dielectric function. To have a better understanding of 
the DFT-GGA method, we adjust the GGA band gap up to the experimental value 
(scissors correction). We find that this adjustment does improve the agreement 
of DFT-GGA with the benchmark methods, but the improvement is not nearly 
enough. Then we calculate the vdW coefficients $C_3$ and $C_5$ for atoms on 
the surface of semiconductors and insulators with the model dynamic multipole 
polarizability and the dielectric functions obtained from the modified Penn 
model, DFT-GGA, $GW$, and $GW$+BSE methods. The results show that, except for 
the vdW coefficients obtained with the DFT-GGA dielectric function, they all 
agree well with each other. The deviations of the vdW coefficients obtained 
with the model dielectric function from those obtained with the $GW$+BSE 
dielectric function are 4\% for $C_3$ and 8\% for $C_5$, respectively. The 
deviation is even smaller between the vdW coefficients obtained from the model 
dielectric function and the $GW$ method. However, these deviations become 
significantly larger for the DFT-GGA ($C_3$: 29\%, $C_5$: 29\%) or scissor-corrected ($C_3$: 24\%, $C_5$: 24\%) dielectric function, 
suggesting the significance of electronic nonlocality that is missing in 
semilocal DFT, leading to the bad performance for the dielectric function of 
ionic solids with strong exciton effect. 

\section{Acknowledgment}
FZ acknowledges support from NSF under Grant no. DMR-1124696. 
JT acknowledges support from NSF under Grant no.  CHE-1261918 and the Office of Naval Research under grant No. N00014-14-1-0761. 
AMR was supported by the Department of Energy Office of Basic Energy Sciences, under Grant no. DE-FG02-07ER15920. 
Computational support was provided by the HPCMO and the NERSC. 

%

\end{document}